\documentclass[12pt]{article}
\usepackage{amssymb,amsmath}
\include{epsf}

\usepackage{graphicx}

\usepackage{amsfonts}

\makeatletter
\@addtoreset{equation}{section} 
\makeatother

\newcommand{\eqn}[1]{(\ref{#1})}

\def\<#1,#2>{\left\langle#1,#2\right\rangle} 

\def\nn{\nonumber}
\newcommand\opname[1]{\mathop{\mathrm{#1}}\nolimits}
\newcommand{\tr}{\mathsf{tr}}
\newcommand{\Tr}{\opname{Tr}}
\newcommand{\del}{\partial}

\newcommand{\dd}{\mathrm{d}}
\newcommand{\ii}{\mathrm{i}}
\newcommand{\e}{\mathrm{e}}
\newcommand{\qj}{\mathsf{j}}
\newcommand{\qf}{\mathsf{f}}

\newcommand{\qT}{\hat{\mathsf{T}}}
\newcommand{\qH}{\mathcal{H}}
\newcommand{\qJ}{\hat{\mathsf{J}}}
\newcommand{\qw}{\mathsf{w}}
\newcommand{\qW}{\hat{\mathsf{W}}}
\newcommand{\qU}{\hat{\mathsf{U}}}
\newcommand{\qV}{\hat{\mathsf{V}}}

\newcommand{\qm}{\mathsf{m}}

\newcommand{\qM}{\hat{\mathsf{M}}}
\newcommand{\qP}{\hat{\mathsf{P}}}

\newcommand{\qX}{\hat{\mathsf{X}}}

\newcommand{\qe}{\mathsf{e}}
\newcommand{\qOmega}{\hat{\mathsf{\Omega}}}
\newcommand{\qGamma}{\hat{\mathsf{\Gamma}}}

\newcommand{\be}{\begin{equation}}
\newcommand{\ee}{\end{equation}}
\newcommand{\bea}{\begin{eqnarray}}
\newcommand{\eea}{\end{eqnarray}}

\begin{document}
\begin{titlepage}
\begin{flushright}
\baselineskip=12pt
DSF/19/2009\\
LPT ORSAY 09.108\\
ICCUB-09-427
\\
December 2009

\end{flushright}

\begin{center}

\baselineskip=24pt

{\Large\bf Star-product in the presence of a monopole}

\baselineskip=14pt

\vspace{1cm}

{\bf Jos\'e F. Cari\~nena$^{a}$, J.M.\ Gracia-Bond\'{\i}a}$^{a}$, {\bf
Fedele Lizzi\,}$^{b,c}$, {\bf Giuseppe Marmo}$^{b}$ and {\bf Patrizia
Vitale}\,$^{b,d}$
\\[6mm] $^a$ {\it Departamento de F\'{\i}sica
Te\'{o}rica, Facultad de Ciencias,
\\
Universidad de Zaragoza, 50009 Zaragoza, Spain}
\\{\tt
jfc@unizar.es,jmgb@unizar.es}\\[6mm]
$^b$ {\it Dipartimento di Scienze Fisiche, Universit\`{a} di Napoli
{\sl
Federico II}
\\
and {\it INFN, Sezione di Napoli, Monte S.~Angelo}\\
Via Cintia, 80126 Napoli, Italy}\\
{\tt fedele.lizzi@na.infn.it, giuseppe.marmo@na.infn.it,
patrizia.vitale@na.infn.it}
\\[6mm]
$^c$ {\it High Energy Physics Group, Dept.\ Estructura i
Constituents de la Mat\`eria, \\Universitat de Barcelona, Diagonal
647, 08028 Barcelona, Catalonia, Spain \\and Institut de Ci\`encies
del Cosmos, UB, Barcelona}
\\[6mm]
$^d$  {\it Laboratoire de Physique Th\'eorique,
Universit\'e Paris XI,
\\
91405 Orsay Cedex, France}

\end{center}

\vskip 2 cm

\begin{abstract}
We present a deformed $\star$-product for a particle in the presence
of a magnetic monopole. The product is obtained within a self-dual
quantization-dequantization scheme, with the correspondence between
classical observables and operators defined with the help of a
quaternionic Hilbert space, following work by Emch and Jadczyk. The
resulting product is well defined for a large class of complex
functions and reproduces (at first order in $\hbar$) the Poisson
structure of the particle in the monopole field. The product is
associative only for quantized monopole charges, thus incorporating
Dirac's quantization requirement.
\end{abstract}

\end{titlepage}

\section{Introduction}

In this paper we perform the deformation quantization for the algebra
of functions on the phase space of an electrically charged particle in
the field of a magnetic monopole at the origin; to be precise, the
phase space is determined by the asymptotics of the 't Hooft--Polyakov
monopole~\cite{DondeBais}. The deformation is therefore that of
complex-valued functions on $\big(\mathbb{R}^3
-\{0\}\big)\times\mathbb{R}^3$ in the presence of the following
Poisson structure: \bea \{p_i,x^j\}&=&\delta^j_i \nonumber\\
\{x^i,x^j\}&=&0
\nonumber\\
\{p_i,p_j\}&=& \frac g2 \varepsilon_{ijk} \frac{x^k}{\vert x\vert ^3},
\label{monopolepoisson}
\eea
where the quantity $g$ is the product of the monopole magnetic
charge and the electric charge of a test particle. Inversion of the
Poisson bracket gives a symplectic two-form
\be
\omega = \dd x^i\wedge \dd p_i + \frac
g2\varepsilon_{ijk}\frac{x^i}{\vert x \vert^3}\dd x^j\wedge \dd x^k.
\label{Poissonmonop}
\ee Existence of a formal $\star$-product is of course ensured by
general theorems for Poisson manifolds~\cite{Kontsevich}, but the aim
of the paper is to follow as close as possible the
Wigner--Weyl--Gr\"onewold--Moyal quantization/dequatization
program~\cite{Weyl}, in order to exhibit the product explicitly. For a
modern treatment of this program, see~\cite{EspositoMarmoSudarshan,
Apophis}. In such a scheme unitary operators are associated to
translations in phase space, while commutation relations reproduce the
exponentiated version of the Poisson structure. In this way not only
the commutation relations of the operators associated to the $x$'s and
$p$'s reproduce (up to~$\ii\hbar$) the corresponding Poisson brackets,
but one is at the same time assured of that the product is well
defined on a large class of functions.

We will therefore seek a generalization of the Weyl map, which
associates operators to functions. To the purpose we use a remarkable
result by Emch and Jadczyk~\cite{EmchJadczyk}, based on quaternionic
quantum mechanics. Here, however, quaternions and quaternion Hilbert
modules are simply a device to build a $\star$-product, which in the
end acts between \emph{complex} functions. It is possible to adapt the
construction for a deformed product between quaternion-valued
functions, but we do not discuss this issue in the present paper.

The paper is organized as follows. In Section~2 we review Weyl systems
and the Weyl--Wigner quantization-dequantization maps for ordinary
quantum mechanics, which brings to the definition of the Moyal
$\star$-product. In Section~3 we give an account of the Emch--Jadczyk
construction, providing a setting for the quantization of a particle
in the field of a magnetic monopole in terms of quaternionic quantum
mechanics. In Section~4 we propose a generalized Weyl system in this
context. Finally, in Section~5 we exhibit a generalized Weyl--Wigner
construction, with a quantization and a dequatization map. This yields
a description of the quantum particle-monopole system in terms of
complex functions, with a new noncommuting $\star$-product. We add
some concluding remarks.

\section{The Weyl--Wigner--Gr\"onewold--Moyal formalism}

Schematically, the quantization of a particle with the usual Poisson
canonical structure on the phase space $T^*\mathbb{R}^3$
(relations~\eqn{monopolepoisson} with~$g=0$) goes as follows. Let us
denote by $\tilde f(\eta,\xi)$ the Fourier transform of the phase-space
function $f(x,p)$. To $f$ it is associated an operator by
performing the inverse transform and inserting in the integral a
unitary operator family $\qT(\eta,\xi)$ instead of $\e^{-\ii(\eta\cdot
p +\xi\cdot x)}$:
\be
\hat f=\hat{\mathcal W}(f) =\frac{1}{(2\pi)^{3}} \int
\dd\eta~ \dd\xi~ \qT(\eta,\xi) \tilde f(\eta,\xi).
\label{weylmap}
\ee
The family $\qT$ is in turn given by $\qT(\eta,\xi):=\e^{\ii(\eta\cdot
\hat P + \xi\cdot \hat X)}$, where $\hat X^i$ and $\hat P_j$ are the
usual position and momentum operators. This Weyl map $f\to\hat f$ is
well defined and invertible for large classes of functions; for
example it associates Hilbert--Schmidt operators to square integrable
functions~\cite{agarwalwolf,pool}. The unitary operators $\qT$ obey
\be
\qT(\eta,\xi)\qT(\eta',\xi') = \qT(\eta+\eta',\xi+\xi')
\e^{\frac\ii2(\eta\cdot\xi'-\eta'\cdot\xi)}.
\label{normalWeyl}
\ee
or, which is the same,
\be
\qT(\eta,\xi)\qT(\eta',\xi')\qT^\dag(\eta,\xi)\qT^\dag(\eta',\xi') =
\e^{\ii(\eta\cdot\xi'-\eta'\cdot\xi)}\hat{I},
\label{wesy}
\ee
and we recognize the canonical symplectic form in the exponential.
They yield a ray representation of the Euclidean group in~3
dimensions, what is called a Weyl system built on the symplectic
vector space $(T{^*R}^3,\omega)$, with the usual canonical commutation
relations between coordinates and momenta descending from~\eqn{wesy}.

The inverse of the Weyl map, defined on suitable domains, is usually
called the Wigner map~\cite{wigner}. It is convenient to express
both the Weyl map and its inverse in terms of two operators,
$\hat\Omega$, $\hat\Gamma$, respectively  called \emph{quantizer}
and \emph{dequantizer}:
\bea
\hat f& =&\hat{\mathcal{W}}(f)=\frac{1}{(2\pi)^3}\int\dd x~ \dd p~ \hat\Omega(x,p)
f(x,p)\nonumber\\
f(x,p)&=&\mathcal{W}^{-1}(\hat f)=\Tr\hat\Gamma(x,p) \hat f.
\eea
For the canonical case, up  to constant factors depending on the normalization conventions,
\be
\hat\Omega(x,p) = \hat\Gamma(x,p) = \int \dd\eta
~\dd\xi~ \qT(\eta,\xi) \e^{-\ii(\eta\cdot p +\xi\cdot x)}.
\label{canonicOmega}
\ee
Notice however that there exist well defined
quantization-dequatization maps for which the two operators are
different~\cite{MMV05,MMMV07}. Functions associated to operators through the
dequatization map, often called symbols, are actually noncommuting:
they reproduce the noncommutativity of the operators by means of a
noncommutative (star) product known as the Gr\"onewold--Moyal
product~\cite{Gronewold, Moyal}:
\be
f\star g=\mathcal{ W}^{-1}\left(\hat{\mathcal{ W}}(f) \hat{\mathcal{
W}}(g)\right).
\ee
This may be written in terms of an integral kernel, in turn
completely specified by the operators $\hat\Omega$ and $\hat\Gamma$
\begin{align}
&f\star g (x,p) = \Tr\hat f\,\hat g\,\hat\Gamma (x,p) = \int \dd
x'\,\dd p'\,\dd x''\,\dd p''\,f(x',p')g(x'',p'') K(x',p';x'',p'';x,p)
\;\mathrm{with}
\notag \\
&K(x',p';x'',p'';x,p) = \Tr\hat\Omega (x',p')\hat\Omega (x'',p'')
\hat\Gamma(x,p) = C \e^{2\ii[x(p'-p'')
+x'(p''-p)+x''(p-p')]}
\label{kernel}
\end{align}
with $C$ a normalization constant.
This associative and noncommutative product is well defined also for
polynomials and reproduces the canonical commutation relations through
the so called Moyal bracket:
\be
x^i\star p_j-p_j\star x^i=\ii\delta^i_j, \quad x^i\star x^j-x^j\star
x^i=p_i\star p_j-p_j\star p_i=0.
\ee
We want to generalize this construction to the particle/monopole
system. We will need therefore the analogue of the operator $\qT$ (a
generalized Weyl system), and a quantization map and a dequatization
map. A generalized Weyl system was proposed by us
in~\cite{proceedings}.

\section{The construction by Emch and Jadczyk}

Quaternionic quantum mechanics is buttressed by the observation that
to describe particles with an inner structure it is necessary to
consider sections of some hermitian complex vector bundle. In this
philosophy, Emch and Jadczyk propose in~\cite{EmchJadczyk} that a
quantum particle in the field of the monopole be described by square
integrable sections of a hermitian quaternionic line bundle over
$\mathbb{R }^3-\{0\}$. In that setting all operators, the generators
of the generalized Weyl system in particular, are quaternionic valued.
As announced earlier, we use their construction as an intermediate
step to build a $\star$-product where operator symbols are genuinely
complex-valued functions. We adopt the following notations.
Quaternionic units are $\qe=(\qe_1,\qe_2,\qe_3)$ and $\qe_0=1$ with
\be
\qe_i\qe_j=-\delta_{ij}\qe_0+\varepsilon_{ijk}\qe_k,~~~i=1,..,3
\ee
and the involution $\qe_0^*=\qe_0,\,\qe_i^*=-\qe_i$. All
complex-valued quantities are in normal typeface: $\alpha,f,\ldots$;
all quaternionic quantities are in~\textsf{sans serif}: $\mathsf{q,
f,j}$; operators are distinguished by the usual hat~$\hat{~}$.
Borrowing the representation of imaginary quaternions by means of
$2\times 2$ skew-hermitian matrices, we may write (when convenient)
\be \qe_0 = \sigma_0, \quad \qe_i = -i\sigma _i, \qquad{\rm and}\qquad
\mathsf{f}(x) = f^0(x)\qe_0 + f^i (x)\qe_i. \ee
Let $\qH$ be the quaternionic Hilbert module formed by sections
$\mathsf\Psi(x)$ with the usual quaternionic-valued inner product
\be
\<\mathsf\Psi,\mathsf\Phi>=\int\dd x~ \mathsf\Psi(x)^*\mathsf\Phi(x)
\ee
and the associated real norm. The elements of the Hilbert module
behave as a vector space under multiplication by quaternions (numbers)
from the \emph{right}, while linear operators act on the \emph{left}.
Among the operators there are of course also the quaternionic valued
functions which act multiplicatively: $\hat\qf\mathsf\Psi(x)
=\qf(x)\mathsf\Psi(x)$. The quaternionic line bundle may be
considered as an associated bundle with structure group $SU(2)$. In
order to lift vector fields from ${\mathbb{R}}^3$ to the total space
of the vector bundle one has to introduce a connection, that is, a
procedure to lift vector fields to horizontal vector fields, which
takes into account the presence of the monopole. This is the {\it
gauge potential}
\be
A = g\,\frac{[{\qe}\cdot{{x}},{\qe}\cdot d{{x}}]}{|{x}|^2},
\label{gaupot}
\ee
with the square bracket indicating the antisymmetrized product and $g$
denoting the product of the electric and magnetic charges. We allow
for the possibility that $g$ be different from 1, which slightly
generalizes the Emch and Jadczyk construction, and will clearly show
how the monopole charge quantization emerges. The origin of the choice
\eqn{gaupot} may be traced back to the Hopf fibration. Since
${\mathbb{S}}^2\times {\mathbb{R}}_+={\mathbb{R}}^3-\{0\}$, we may
define a lifting which would consider wave functions as fields
transforming covariantly under the rotation group, whose action in the
inner space is by means of~$SU(2)$. Given any ${{u}}\in
{\mathbb{S}}^2$, the translation ${{u}}\cdot{\partial}/
{\partial{{x}}}$ lifts to the quaternionic-valued differential
operator
\be
\nabla_{u}= \qe_0{u}\cdot\frac{\partial}{\partial x} + \frac{g}{2}
\frac{[ \qe\cdot x, \qe\cdot u]} {|x|^2} = \qe_0
u\cdot\frac{\partial}{\partial x} + \frac{g}{2}\,\qe\cdot\frac{u\wedge
x}{|x|^2}.
\label{tra}
\ee
We use the notation $\nabla_i$ for the covariant derivatives along the
cartesian basis vectors. It may be verified that it obeys
$[\nabla_i,x_j]=\delta_{ij}\,\qe_0$; thus it generates translations on
configuration space. Moreover it transforms as a vector under
rotations: $[M_i,\nabla_j]=-\varepsilon_{ijk}\nabla_k$, with $M_i=
\varepsilon_{ijk}x_j\del_k-\qe_i/2$. Central to the construction is
the following imaginary unit introduced by Emch and Jadczyk:
\be
\qj(x)=\frac{\qe\cdot x}{|x|}; \quad \qj^2=-\qe_0, \quad \qj^*=-\qj.
\ee
It is also rotationally invariant (we regard the $\qe$'s as
transforming like the components of a vector under rotations).
Associated with $\qj$, consider the linear operator $ \qJ$,
commuting with translations:
\be
\qJ\mathsf\Psi(x) := \qj(x)\mathsf\Psi(x).
\ee
Infinitesimal translations do not commute among themselves, their
commutator being the monopole field
\be
[\nabla_i,\nabla_j]=-\frac{1}{2}g\varepsilon_{ijk}\frac{x^k}{|x|^3}
 \qJ.
\ee
Finite translations $\qU(\eta)$ generated by the operator
$\nabla_\eta$, for $\eta\in \mathbb{R}^3-\{0\}$, acquire a quaternionic
phase factor. We have indeed
\be
\qU(\eta)\mathsf\Psi(x) = \qw(\eta,x-\eta)\mathsf\Psi(x-\eta),
\label{U}
\ee
where $\qw(\eta,x)$, for every $\eta$ not collinear with $x$, is
given by the quaternion
\be
\qw(\eta,x)=\exp\left( \qj(x\wedge \eta) \frac{g\alpha}{2}\right),
\ee
with $\alpha$ the angle between $x$ and $x+\eta$. This may be
checked directly, on the strength of
\be
\nabla_{ \eta} \mathsf\Psi(x) = \lim_{t\rightarrow
0}\left(\frac{d}{dt} (\qU(t \eta) \mathsf\Psi)(x)\right)
\;\mathrm{and}\;
\lim_{t\rightarrow 0}\frac{\dd}{\dd t}\qw(t\eta,x-t\eta)=
\frac{g}{2}\,\qe \cdot \frac {x \wedge \eta}{\vert x\vert ^2}.
\label{useful}
\ee
Some easily verifiable properties of $\qw$, satisfied almost
everywhere, \label{pagetest} are:
\begin{enumerate}
\item
$\qw(0,x)= 1$ \label{prop1}
\item
$\qw(\eta,x) \qw^*(\eta,x)=1$ \label{prop2}
\item
$\qw^*(\eta, x)= \qw(-\eta,x+\eta)$
\item
$\qw(t  \eta,   x+s  \eta)\qw(s  \eta, x)=\qw((s+t)\eta,x)$.
\end{enumerate}
with $t,s$ real parameters. We also consider the multiplication
operators $\qW(\eta)\mathsf\Psi(x):=\qw(\eta,x)\mathsf\Psi(x)$.
Operators $\qU(\eta)$ may be written as the product of ordinary
translations in $\mathbb{R}^3$ and the $\qW(\eta)$:
\be
\qU(\eta)=\qV(\eta)\qW(\eta)
\quad \mathrm{with} \quad
\qV(\eta)\mathsf\Psi(x) = \mathsf\Psi(x-\eta).
\ee
The operator $\nabla$ is skew-hermitian. Let us define the hermitian
operators
\be
\qP_i = -\qJ\nabla_i = -\nabla_i\qJ.
\ee
We have
\be
[\qP_i,\qP_j]=\frac12\varepsilon_{ijk} g \frac{x^k }{ |x|}\qJ.
\label{pp}
\ee
The $\qP_i$ are generators of translations in the quaternionic Hilbert
space. Notice that the two summands in $\qP_i$ do not commute. Finally,
\begin{equation}
\qU({\eta}) \qU({\eta'}) \mathsf\Psi({x}) = \qw({\eta};{x}-{\eta})
\qw({\eta'};{ x}-{\eta}-{\eta'}) \mathsf\Psi({x}-{\eta}-{\eta'}).
\label{uaub}
\end{equation}
The key result of Emch and Jadczyk is that this may be written in
terms of $\qU({\eta}+{\eta'})$:
\begin{equation}
\qU({\eta}) \qU({\eta'})=\qU({\eta} + {\eta'}) \qM({\eta}, {\eta'}).
\label{transprod}
\end{equation}
by means of multiplication operator,
\be
\qM({\eta}, {\eta'}) \mathsf\Psi({ x}) := \mathsf{m}({\eta}, {\eta'};
x)\mathsf\Psi( x)
\ee
with $\mathsf{m}({\eta},{\eta'};{ x}) = \qw^*({\eta} + {\eta'}, { x}
)\qw({\eta}, { x} +{\eta'}) \qw( {\eta'}, { x} )$.
Since $\qw({0},{ x})=1$ and $\qw(\eta,{  x -\eta})=\qw^*({-\eta}, {
x})$ we have that
\begin{equation}
\mathsf{ m}({ \eta};-{ \eta}; x)=1 \label{prom}.
\end{equation}
The quantity $\qm$ can be expressed in exponential
form~\cite{EmchJadczyk}:
\be
\qm(\eta,\eta';x)=\exp\left( \frac{g}{4}\qj(x)\epsilon_{i j k}
\frac{x^i}{|x|^3} \eta^j \eta'^k\right) \label{m}
\ee
where we recognize the flux of the monopole field through the flat
triangle with vertices $(x, x+\eta,x+\eta+\eta')$. This result may be
easily obtained by direct calculation, observing that~\eqn{transprod}
implies
\begin{equation}
\qM({\eta}, {\eta'}) =\qU^{-1} ({\eta} + {\eta'}) \qU({\eta})
\qU({\eta'})
\label{mop}
\end{equation}
with $\qU(\eta)$ in the form $\qU(\eta)=\exp(\qJ\eta\cdot\qP)$ and
repeteadly using the commutation relation~\eqn{pp}. The associativity
condition $\big(\qU(\eta)\qU(\eta')\big)\qU(\eta'')=
\qU(\eta)\big(\qU(\eta')\qU(\eta'')\big)$ requires
$$
\qM(\eta,\eta')\qM(\eta+\eta',\eta'') = \qM(\eta,\eta'+\eta'')
\qM(\eta',\eta'').
$$
This last equality is satisfied only if the flux through the
tetrahedron identified by the vectors
$x,x+\eta,x+\eta+\eta',x+\eta+\eta'+\eta''$ is a multiple integer od
$2\pi$, that is only if $g$ is an integer. This is the celebrated
\emph{quantization condition}. We see that the construction will yield
an associative algebra only for systems which respect it. Hereafter
for simplicity of notation we will consider the case $g=1$.

Summarizing, the Emch--Jadczyk here reviewed provides a unitary
representation for noncommuting translations in terms of quaternionic
valued operators acting on quaternionic wave functions. This is the
first stone of our construction.

\section{The monopole Weyl system}

On the quaternionic Hilbert module $\qH$, we consider the six
operators
\bea
\qX^i\mathsf\Psi(x)&=&x^i\mathsf\Psi(x),
\nonumber \\
\qP_j\mathsf\Psi(x)&=& -(\qJ \nabla_i  \mathsf\Psi)(x),
\eea
with commutation relations
\be
\begin{array}{rcl}
[\qP_i, \qX_j] &=& -\qJ \delta_{ij},\nonumber
\\
{[} \qP_i, \qP_j{]} &=& \frac 12  g \qJ \, \varepsilon_{ijk}
{\displaystyle\frac{x^k}{| {x}|^3}},
\\
{[}\qX_i, \qX_j{]} &=& 0.
\nonumber
\end{array}
\ee
They may be regarded as a deformation of the Euclidean algebra in~3
dimensions, although do not define a Lie algebra anymore. It is
however possible to exhibit a unitary representation of the relations,
what we call a generalized Weyl system~\cite{proceedings}. This is
provided by the operator family
\begin{equation}
\qT(\alpha) = \e^{\qJ[ { \eta}\cdot \qP+ { \xi}\cdot\qX]}=
\e^{\qJ { \eta}\cdot {\qP}} \e^{\qJ
  { \xi}\cdot\qX}  \e^{\frac{1}{2}\eta^i \xi^{j}
[\qP^i,\qX^j]}=\e^{\qJ { \eta}\cdot{\qP}} \e^{\qJ { \xi}
\cdot{\qX}}\e^{-\frac{1}{2}\qJ { \eta}\cdot { \xi}} =\e^{\qJ {
\xi}\cdot{\qX}} \e^{\qJ { \eta}\cdot {\qP}} \e^{\frac{1}{2}\qJ {
\eta} \cdot { \xi}}\label{pqexp},
\end{equation}
with $\alpha=({ \eta},{ \xi})$. Remember that $ \exp ( \qJ {
\eta}\cdot {\qP}) \equiv \qU({ \eta})$. We have then
\begin{equation}
\qT(\alpha)\mathsf\Psi(x) = \e^{\qJ \eta\cdot \qP} \e^{\qJ \xi \cdot
\qX} \e^{-\frac{1}{2}\qJ \eta\cdot \xi} \mathsf\Psi({x})=\qw({
\eta};{x -\eta})\e^{\qj({x -\eta}) { \xi}\cdot({x
-\eta})}\e^{-\frac{1}{2}\qj({x -\eta}) { \eta} \cdot
{\xi}}\mathsf\Psi({x -\eta}),
\end{equation}
but also
\begin{equation}
\qT(\alpha) \mathsf\Psi({x} = \e^{\qJ { \xi} \cdot{\qX}} \e^{\qJ {
\eta}\cdot {\qP}}\e^{\frac{1}{2}\qJ { \eta} \cdot {
\xi}}\mathsf\Psi({x}) = \e^{\qj({x}) {\xi\cdot x}}\qw({ \eta}; {x
-\eta})\e^{\frac{1}{2}\qj({x -\eta}) { \eta} \cdot {
\xi}}\mathsf\Psi({x -\eta})\,.
\end{equation}
We compute
\begin{equation}
\qT(\alpha) \qT(\beta)=\e^{\qJ[{ \eta}\cdot {\qP}+ {
\xi}\cdot{\qX}]}\e^{\qJ[ { \eta'} \cdot{\qP}+ {
\xi}'\cdot{\qX}]}=\e^{\qJ { \eta}\cdot {\qP}} \e^{\qJ {
\xi}\cdot{\qX}} \e^{\qJ { \eta'}\cdot
{\qP}} \e^{\qJ { \xi}'\cdot{\qX}}\e^{-\frac{1}{2}\qJ({ \eta} \cdot {
\xi}+{ \eta'} \cdot { \xi}')}\,.
\end{equation}
On using~(\ref{transprod}) and~(\ref{pqexp}), we arrive at the sought
for generalization of~\eqn{wesy}:
\begin{equation}
\qT(\alpha) \qT(\beta)=\qT(\alpha+\beta)\qM({ \eta},{ \eta'})
\exp\Bigl(\frac{1}{2}\qJ ({ \eta}\cdot { \xi}'-{ \eta'}\cdot
{ \xi})\Bigr)\,. \label{weylsys}
\end{equation}
Like its usual counterpart~\eqn{normalWeyl}, this generalized Weyl
system provides a projective representation of the translation group,
but in this case there are two phases. One is present also in the
usual quantization scheme and gives the noncommutativity of positions
and momenta ---here however with the imaginary unit replaced by the
quaternionic radial function~$\qj({x})$. The factor~$\qM$ instead
contains the information on the noncommutativity of the translations.

On using the identity
\be
{1}= \qT(-\beta)\qT(-\alpha)\qT(\alpha)
\qT(\beta)=\qT(-\alpha-\beta)\qM({- \eta'},{-
\eta})\qT(\alpha+\beta)\qM({ \eta},{ \eta'}),
\ee
and observing that, from~\eqn{weylsys}
\begin{align}
&\qT(-\alpha)\qT(-\beta)\qT(\alpha)
\qT(\beta)
\notag \\
&= \qT(-\alpha-\beta)\qM({- \eta},{- \eta'})\qT(\alpha+\beta)\qM({
\eta},{ \eta'}) \exp\Bigl(\qJ ({ \eta}\cdot { \xi}'-{ \eta'}\cdot {
\xi})\Bigr),
\end{align}
we may rewrite the generalized Weyl system as
\be
\qT(-\alpha)\qT(-\beta)\qT(\alpha) \qT(\beta)=\qM({ \eta'},{
\eta})^{-1}\qM({ \eta},{ \eta'}) \exp\Bigl(\qJ ({ \eta}\cdot {
\xi}'-{ \eta'}\cdot { \xi})\Bigr),
\ee
similar to~\eqn{wesy}.

\section{Monopole quantization/dequantization maps}

Next we exhibit generalized Weyl and Wigner maps, associating to
the classical observables operators on the quaternionic Hilbert module
and viceversa; then, following the deformation quantization programme,
we introduce a noncommuting $\star$-product for the algebra of
operator symbols (complex-valued functions). To a complex function on
phase space $f(x,p)=f_r(x,p)+\ii f_i(x,p)$, where $f_r$ and $f_i$ are
real, let us associate a quaternion by just substituting the
quaternionic coordinate-dependent unit $\qj(x)$ for the imaginary
unity $\ii$; so that
\be
f\longrightarrow \qf(x,p)=f_r(x,p)\qe_0 +\qj(x) f_i(x,p).
\ee
This map is obviously invertible with inverse map
\be
f(x,p)=\frac12[\tr\, \qf -\ii\,\tr\,(\qj(x) \qf)];
\ee
Here $\tr$ is the quaternionic trace: with quaternions represented as
$2\times 2$ Pauli matrices, $\tr\qe_0=2$ and $\tr\qe_i=0$. The
quaternionic-valued operator
\be
\hat\qf=\frac{1}{(2\pi)^3} \int \dd x~ \dd p~
\dd\eta~ \dd\xi~ \e^{-\qJ(\xi x + \eta p)} \e^{\qJ(\xi \qX + \eta
\qP)} (f_r(x,p)+\qJ f_i(x,p))\equiv \hat\qf_r + \qJ \hat\qf_i
\label{qmap}
\ee
is of the form
\bea
\hat\qf_{r,i} &=& \int \dd x~ \dd p~ f_{r,i}(x,p)\qOmega(x,p),
\eea
where we read the \emph{quantizer} $\qOmega$ off~\eqn{qmap}:
\be
\qOmega (x,p)=\int \dd \eta ~\dd\xi~ \e^{-\qJ (\xi\cdot x +
\eta\cdot p)}\qT(\alpha) = \int \dd \eta~ \dd\xi~ \e^{-\qJ (\xi\cdot
x + \eta\cdot p)} \e^{\qJ[ {\eta}\cdot \qP+ {\xi}\cdot\qX]}.
\label{quantizer}
\ee
This is to be compared with the canonical one \eqn{canonicOmega}.

Now we claim that the dequantization map is given by:
\bea
f(x,p)&=&\frac12\tr\int\dd\xi~ \dd\eta~ \Tr_{op} \e^{\qJ(\xi x +
\eta p)} \e^{-\qJ(\xi \qX + \eta \hat{ P})}\hat\qf \nonumber\\&&
-\frac\ii2 \tr \int\dd\xi~ \dd\eta~ \Tr_{op} \e^{\qJ(\xi \cdot x +
\eta \cdot p)} \e^{-\qJ(\xi \qX + \eta \hat{ P})}\qJ\hat\qf,
\label{inverse}
\eea
conveniently rewritten as
\be
f(x,p)=\frac12\tr \Tr_{op} \hat\qf \qGamma(x,p)-\frac\ii2 \tr
\Tr_{op} \qJ\hat\qf\qGamma(x,p) \label{deqmap},
\ee
with the quantization being self-dual in that the dequantizer actually
is equal to the quantizer:
\be
\qGamma(x,p)=\qOmega(x,p).
\ee
We need to prove that \eqn{inverse} is indeed the inverse of
\eqn{qmap}. This amounts to show that
\bea
\tr\Tr_{op} \qOmega(x,p) \qGamma(x',p')& =& \delta(x-x')
\delta(p-p')\label{finv}\\
\tr\Tr_{op} \qJ \qOmega(x,p) \qGamma(x',p') &=&0
 \label{inverseJ}
\eea
For this we refer to Appendix~\ref{se:inverJ}.

\subsection{The $\star$-product}

We can now proceed to define the star product as in~\eqn{kernel}
\be
f\star g (x,p)= \frac12\left(\tr\Tr_{op} \hat\qf
\hat{\mathsf{g}} \qGamma (x,p)
-\ii \tr\Tr_{op}\qJ  \hat\qf \hat{\mathsf{g}} \qGamma (x,p)\right),
\ee
As for the integral kernel, we define, analogously to~\eqn{kernel}
\begin{align*}
K_1(x',p';x'',p'';x,p) &= \frac12 \tr \Tr_{op} \qOmega (x',p')\qOmega
(x'',p'')\qGamma(x,p).
\\
K_2(x',p';x'',p'';x,p) &= \frac12\tr \Tr_{op} \qJ\qOmega
(x',p')\qOmega (x'',p'')
\qGamma(x,p).
\\
K_M(x',p';x'',p'';x,p) &= K_1(x',p';x'',p'';x,p) - \ii
K_2(x',p';x'',p'';x,p)
\label{kernelquat}.
\end{align*}
With this notation, the star product acquires the usual integral
kernel form
\be
f\star g (x,p)= \int\dd x'\dd p' \dd x'' \dd p''
~ f(x',p') g(x'',q'') K_M(x',p';x'',p'';x,p)
\ee
The explicit expression of $K_M$ can be calculated observing that
\begin{align}
K_1(x',p';x'',p'';x,p) &= \tr\bigl[\exp\bigl(2\qj(x+x'-x'')((x'-x)
p''+(x''-x') p+ (x-x'') p)\bigr)
\nn \\
&\quad \qm\left( 2(x'-x), 2(x''-x'); x-x''+x'\right)\bigr]
\nn \\
K_2(x',p';x'',p'';x,p) &= \tr\bigl[\exp\bigl(2\qj(x+x'-x'')((x'-x)
p''+(x''-x') p+ (x-x'') p)\bigr)
\nn \\
&\quad \qm\left( 2(x'-x), 2(x''-x'); x-x''+x'\right)\qj(x+x'-x'')\bigr]
\end{align}
with $\qm\left( 2(x'-x), 2(x''-x'); x-x''+x'\right)$from \eqn{m}, given by
\be
\qm=\exp \left(\frac {g}{|x+x'-x''|^3}
\qj(x+x'-x'')\epsilon_{ijk}(x+x'-x'')^i (x'-x)^j (x''-x')^k\right).
\ee
We have then for $K_M(x',p';x'',p'';x,p)$ the expression
\be
K_M=C\exp\left[2i\left((x'-x)\cdot p''+(x''-x')\cdot p+ (x-x'')\cdot
p' +\frac{g}{2} \frac{x \cdot(x'\wedge x'')}{|x-x''+x'|^3}
\right)\right]
\ee
and $C$ is a normalization constant.
In the case $g=0$ we recover the result for the Moyal
kernel~\eqn{kernel}.

We may use this result to compute the star product of the coordinate
functions in phase space. Products which involve at least one
coordinate function $x^i$ are easy to calculate, less trivial is the
computation of the star product of momenta. We find:
\bea
x^i\star x^j&=&x^i x^j\\
x^i\star p_j&=& x^i p_j - \frac{\ii}{2} \delta^i_j\\
p_i\star p_j &=& p_i p_j -\frac{\ii}{4} g
\epsilon_{ijk}\frac{x^k}{|x|^3}
\eea
This obviously reproduces the Poisson structure.

\section{Concluding remarks}
The conjecture of the authors in~\cite{proceedings}, that a
Weyl--Wigner--Groenewold--Moyal quantization of a phase space in the
(asymptotic) type of a 't Hooft--Polyakov monopole can be effected, by
choosing as fundamental complex structure the quaternionic operator
$\qj$ studied by Emch and Jadczyk, has been proved.

\subsection*{Acknowledgments} JMG-B is grateful to the Dipartimento di
Scienze Fisiche, Universit\`{a} di Napoli, for customary
hospitality. His work was supported as well by DGIID-DGA (grant
2008-E24/2). FL~would like to thank the Department of Estructura i
Constituents de la materia, and the Institut de Ci\`encies del
Cosmos, Universitat de Barcelona for hospitality. His work has been
supported in part by CUR Generalitat de Catalunya under project
2009SGR502.  P. V. thanks the LPT Orsay for hospitality. Her work
was supported in part by the European Science Foundation Research
Networking Program ``Quantum Geometry and Quantum Gravity''
(Exchange Visit Grant 2595).

\appendix

\setcounter{section}{0}
\renewcommand{\thesection}{\Alph{section}}

\section{Proof of  \eqn{finv}, \eqn{inverseJ} \label{se:inverJ}}
As for \eqn{finv}
\bea
&& \tr\Tr_{op}
\qOmega(x,p) \qGamma(x',p') =\int \dd\eta~ \dd\xi ~ \dd\eta' ~ \dd\xi'\tr\Tr_{op}
\qOmega(x,p) \qGamma(x',p')\nn \\
&=&\int \dd\eta~\dd\xi ~ \dd\eta' ~ \dd\xi'\tr\Tr_{op}
\left[\qT(\alpha)  \qT(\beta) \exp[-\qJ(\xi x+\eta
p)]\exp[-\qJ(\xi'x' +\eta' p')]\right]\nn\\
&=& \int \dd\eta~ \dd\xi ~ \dd\eta' ~ \dd\xi'\tr\Tr_{op}\left[
\qT(\alpha+\beta) M(\eta,\eta')
e^{\frac{1}{2}\qJ(\eta \xi'-\eta' \xi)} \exp[-\qJ(\xi x+\eta
p)]\exp[-\qJ(\xi'x +\eta' p')]\right] \nn\\
&=&\int \dd\eta~ \dd\xi ~ \dd\eta' ~ \dd\xi' \tr\Tr_{op}
\left[\e^{\qJ(\eta +\eta')\qP }e^{J(\xi +\xi')\qX}
e^{-(\eta+\eta')(\xi+\xi')/2}\qM(\eta,\eta')\right.\nn\\
&& \left. \e^{\frac{1}{2}\qJ(\eta \xi'-\eta'\xi)} \exp[-\qJ(\xi
x+\eta p)]\exp[-\qJ(\xi'x' +\eta' p')]\right] \nn\\
&=&\int \dd\eta~ \dd\xi ~\dd\eta' ~\dd\xi' ~\dd y\tr\e^{-(\xi
x+\eta p+\xi'x'+\eta' p')\qj(y)} \delta(\eta+\eta') \qw(\eta+\eta',
y-\eta-\eta')\nn\\
 && \cdot \e^{\qj(y)(\xi +\xi')y}\e^{\qj(y)(\xi
+\xi')(\eta+\eta')}\mathsf{m}(\eta,\eta' ;y) \e^{\qj(y)(\eta
\xi'-\eta'\xi)/2}\nn\\
 &=&\int \dd\xi  \dd\eta \dd\xi' \dd y \tr \e^{-(\xi
x+\xi'x'+\eta(p -p'))\qj(y)}  \e^{j(y)(\xi +\xi')y}
\e^{\qj(y)\eta(\xi +\xi')/2}\nn\\
 &=&\int \dd\xi  ~\dd\xi'~ \dd y ~\tr\e^{-(\xi
x+\xi'x')\qj(y)}  \e^{\qj(y)(\xi +\xi')y} \delta( \frac{\xi'+\xi
}{2}-(p-p')) \nn\\
 &=&\int \dd\xi   \dd y\tr \e^{-[(
\xi -x') x+2(p-p')x']\qj(y)}  \e^{j(y)2(p-p') y}  \nn\\
 &=&\int    \dd y~\tr \e^{-2(p-p')x' \qj(y)}
\e^{\qj(y)2(p-p') y} \delta(x-x') =\int \dd y~
\tr\e^{\qj(y)2(p-p')(x'-y)}\delta(x-x')\nn\\
  &=& 2\int \dd y ~\cos[2(p-p')(x'-y)]\delta(x-x')=
\delta(p-p') \delta(x-x').
\eea
The proof of \eqn{inverseJ} is similar.

\end{document}